\documentclass[%
 reprint,
 superscriptaddress,
 amsmath,amssymb,
 aps,
pra,
showpacs]{revtex4-1}

\usepackage{graphicx}
\usepackage{dcolumn}
\usepackage{bm}

\newcommand{\bra}[1]{\langle #1 \vert}
\newcommand{\ket}[1]{\vert #1 \rangle}
\newcommand{\eq}{Eq.~}
\newcommand{\eqs}{Eqs.~}
\newcommand{\fig}{Fig.~}
\newcommand{\figs}{Figs.~}
\newcommand{\cf} {cf.~}
\newcommand{\ug} {=}
\newcommand{\piu} {+}
\newcommand{\meno} {-}

\newcommand{\eg} {e.g.~}
\newcommand{\rref} {Ref.~}

\begin{document}

\preprint{APS/123-QED}

\title{Quantum-state transfer in staggered coupled-cavity arrays}

\author{Guilherme M. A. Almeida}
\email{gmaalmeidaphys@gmail.com}
\affiliation{%
 Departamento de F\'{i}sica, Universidade Federal de Sergipe, 49100-000 S\~{a}o Crist\'{o}v\~{a}o, Brazil
}%

\author{Francesco Ciccarello}%
\affiliation{NEST, Istituto Nanoscienze-CNR and Dipartimento di Fisica e Chimica, Universit$\grave{a}$  degli Studi di Palermo, via Archirafi 36, I-90123 Palermo, Italy}

\author{Tony J. G. Apollaro}%
\affiliation{NEST, Istituto Nanoscienze-CNR and Dipartimento di Fisica e Chimica, Universit$\grave{a}$  degli Studi di Palermo, via Archirafi 36, I-90123 Palermo, Italy}
\affiliation{Centre for Theoretical Atomic, Molecular, and Optical Physics, School of Mathematics and Physics, Queen's University Belfast, BT7,1NN, United Kingdom}

\author{Andre M. C. Souza}%
\affiliation{%
 Departamento de F\'{i}sica, Universidade Federal de Sergipe, 49100-000 S\~{a}o Crist\'{o}v\~{a}o, Brazil
}%

\date{\today}

\begin{abstract}

We consider a coupled-cavity array, where each cavity interacts with an atom under 
the rotating-wave approximation. For a staggered pattern of inter-cavity couplings, a 
pair of field normal modes each bi-localized at the two array ends arise. A rich structure of dynamical regimes can hence be addressed depending on which resonance condition between the atom and field modes is set.
We show that this can be harnessed to carry out high-fidelity quantum-state transfer (QST) of photonic, atomic or polaritonic states. Moreover, by partitioning 
the array into coupled modules of smaller length, the QST time can be substantially shortened without significantly affecting the fidelity.

\end{abstract}

\pacs{03.67.Hk, 42.50.Pq}
  
  \maketitle


\section{\label{sec1}Introduction}

The potential of coupled high-quality cavities as a platform for simulating many-body quantum phenomena 
has attracted considerable interest over the past few years \cite{hartmann08rev, tomadin10}. 
Such an architecture would indeed enable a high degree of control and addressability of individual sites. Moreover, the coupling to atoms 
results in the formation of polaritons (pseudo-particles 
involving atomic and photonic excitations), which
can give rise to novel strongly correlated regimes of light and matter.

A prototype of such systems is a coupled-cavity array (CCA) described by the so called 
Jaynes-Cummings-Hubbard (JCH) model \cite{angelakis07, greentree06}, where -- due to the overlap 
between evanescent field modes -- photons can hop across nearest-neighbour cavities and at the same time interact 
with two-level quantum emitters (``atoms"). 
In the strong atom-field coupling regime, an effective repulsive 
photon-photon interaction takes place resulting in a Mott-insulator
state for the system \cite{angelakis07, greentree06, hartmann06, rossini07, aichhorn08, koch09, schmidt09, pippan09, schmidt10, knap10}. 
The competition between this photon-blockade effect \cite{birnbaum05} 
and the photon hopping creates a Mott-insulator--superfluid 
quantum phase transition in analogy with the Bose-Hubbard model \cite{fisher89}.

Besides being promising quantum simulators (\cf \rref\cite{raftery14} 
for a recent implementation of a Jaynes-Cummings dimer in a superconducting circuit), 
coupled-cavity networks are attractive platforms for distributed quantum information 
processing and quantum communication \cite{kimble08, cirac97, serafini06}. 
Among its crucial requirements, a quantum network must be capable of
creating entanglement, performing quantum gates and transmitting quantum states
between arbitrarily distant nodes. As atomic systems are long-lived quantum 
memories and photons can faithfully carry information over long distances, 
hybrid atom-photon interfaces indeed appear to be ideal building blocks
of a quantum network architecture 
\cite{chaneliere05,ritter12}. 

From this perspective, a key issue is the study of excitation transport -- in the form of photonic, atomic or polaritonic excitations -- as well as quantum-state transfer (QST) \cite{bose03,apollarorev} across CCAs \cite{nohama07, bose07-qst, hu07, ogden08, makin09, lu10, ciccarello11, dong12, dong13, almeida13, quach13, Latmiraletal15, behzadi13}. 
Non-trivial dynamics are also exhibited by CCAs featuring only a single cavity coupled to an atom
\cite{zhou08, longo10, biondi14, felicetti14, lombardo14,tiecke14}.

In this paper, we explore the potential of a CCA to work as a bus for achieving high-fidelity QST without demanding any dynamical control or measurement.
QST is a pivotal task in quantum communication, which has been intensively investigated mostly in connection with 
spin chains following the seminal proposal by Bose \cite{bose03} (for a review see \eg \rref\cite{apollarorev}). Given 
an array of coupled qubits (such as a spin chain), the goal of QST is transferring an arbitrary quantum state of a qubit located at one end of the array to the qubit at the opposite end. This
should be performed by simply letting the many-qubit system to evolve in time according to its Hamiltonian.
Achieving this with high-efficiency is, in general, non-trivial. For instance, this is not possible in chains (especially long ones) with uniform spin-spin couplings \cite{bose03} due to the detrimental dispersion of the initial wave packet. To get around it, several schemes were thus put forward. It was shown, in particular, 
that perfect length-independent QST can be reached by engineering the spin-spin couplings so as to induce a linear dispersion relation \cite{christandl04, difranco08} (see also Ref. \cite{plenio04} for coupled harmonic systems). 
This yields a ballistic QST, entailing that the QST time is proportional to the chain length.
A reliable local modulation involving the entire chain, however, would face several practical difficulties on the experimental side.  Ballistic QST can also be achieved under appropriate tuning of the outermost couplings
\cite{apollaro12, zwick12}.
A different approach relies on the weak interaction of the sender and receiver
spins with a bulk embodied by a uniform chain \cite{wojcik05, venuti07-2}. 
Schemes of this kind exploit the appearance of a pair of Hamiltonian eigenstates strongly bi-localized at the outermost
weakly-coupled sites (behaving as chain defects), which brings about an effective Rabi-like dynamics \cite{wojcik05}. 
A similar dynamics can be triggered by applying strong magnetic fields on the sender and receiver qubits or their nearest neighbors \cite{plastinaPRL2007, lorenzo13, paganelli13}.
At variance with ballistic QST protocols, a usual drawback of Rabi-like mechanisms is that they typically require long QST times.
%

Here, we assume a scheme of staggered inter-cavity coupling strengths, also 
known as the Peierls distorted chain \cite{peierlsbook}, 
which has been addressed for QST \cite{huo08,kuznetsova08} and quantum teleportation protocols~\cite{venuti07, giampaolo09, giampaolo10} in spin systems (CCAs were considered for implementing distorted chains in Refs.~\cite{giampaolo09, giampaolo10}).
This model also belongs to the class of QST schemes relying on Rabi-like dynamics, hence requiring relatively long transfer times. One of our goals is to keep a high-quality QST via Rabi-like dynamics but, at the same time, significantly reduce the required transfer time. 
We show that this can be achieved by {\textit{modularizing}} the array, namely
connecting identical subunits of Peierls distorted chains.  We first discuss this in detail for
an atom-free CCA, which also applies to any spin chain (irrespective of its realization) having an analogous pattern of couplings.
We then show how to exploit these findings when the CCA is coupled to atoms in order to devise
schemes for transferring atomic or polaritonic states.
%

The present paper is organized as follows. 
In Section \ref{sec2}, we study the single-photon spectrum and stationary states of
a staggered CCA, highlighting in particular the features that are crucial for QST purposes. In Section \ref{review},
we review the basic ideas of QST in spin chains with a special focus on those schemes
whose working principle relies on the formation of bi-localized states. In Section \ref{atomfree}, we study
QST across a staggered atom-free CCA. In Section \ref{modularization}, we show how
the staggered CCA can be modified so as to shorten
the QST transfer time. In Section \ref{sec3}, we study the CCA dynamics in the presence
of atoms and the regimes that are relevant for QST. In Section \ref{sec5}, we show how
to achieve QST of atomic and polaritonic qubits. Finally, in Section \ref{sec6} we draw our conclusions.

\section{\label{sec2}CCA with staggered hopping rates}

\begin{figure}[t] 
\includegraphics[width=0.48\textwidth]{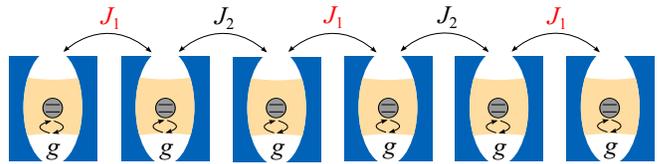}
\caption{\label{fig1} (Color online) Sketch of a CCA with a staggered pattern of hopping rates, where 
$J_1   \ug (1\piu \eta) J$ and $J_2  \ug (1\meno\eta) J$. The protected mode of each cavity can be coupled to a two-level atom with rate $g$.}
\end{figure}

Our set-up consists of a CCA comprising an {\it even} number $N$ of identical, single-mode, lossless cavities. 
Nearest-neighbour cavities are coupled according to a {\it staggered} pattern of hopping rates such that
two possible hopping rates $J_1 $ and $J_2 $ are interspersed along the array, as sketched in \fig\ref{fig1}.
Each cavity in turn (see \fig\ref{fig1}) can be coupled to a two-level quantum emitter (atom). 

In this and the following three sections, we shall focus on the {\it free} field Hamiltonian, i.e., that of an
atom-free CCA. We will consider the full setup, including the atoms, starting from Section \ref{sec3}.

The free field Hamiltonian of the staggered CCA is modelled as (we set $\hbar \ug1$ throughout)
\begin{equation} \label{Hf}
\hat H_{\mathrm{hop}}\ug - J\!\sum_{x=1}^{N-1}[1-(-1)^{x} \eta] \,(\hat a_{x+1}^{\dagger}\hat a_{x}+\mathrm{H.c.}),
\end{equation}
where the bosonic ladder operator $\hat a_{x}^{\dagger} $ ($\hat a_{x}$) creates (annihilates) a photon at the $x$th
cavity.  Note that for odd (even) $x$ the quantity between square bracket in \eq(\ref{Hf}) equals $J_1   \ug(1\piu\eta) J$ [$J_2  \ug (1\meno\eta) J$], where $J$ 
sets the hopping scale
and $-1\!\le\!\eta\!\le\!1$ is a dimensionless distortion parameter (rates will be always expressed in units of $J$). 
For $\eta\ug0$, we retrieve the CCA with uniform inter-cavity couplings usually considered in JCH models \cite{makin09}.
We also point out that, since $N$ is even, for 
$\eta\rightarrow -1^+$ 
the two outermost
cavities (corresponding to $x\ug1$ and $x\ug N$, respectively) are weakly coupled to the remaining ones (bulk), a property which will be crucial for our goals. In assuming that the free field Hamiltonian 
is given by \eq(\ref{Hf}), we have neglected the usual on-site contribution $\sum_x \omega_c\, \hat a_x^\dag\hat a_x$ with $\omega_c$ being the frequency of the each cavity protected mode, which is equivalent
to set the energy scale such that $\omega_c\ug0$.

Our first task is to diagonalize Hamiltonian (\ref{Hf}) 
in the single-photon Hilbert space, which is spanned by the basis $\{|x\rangle\}$ with $|x\rangle\ug \hat a_x^\dag|{\rm vac}\rangle$ 
and $|{\rm vac}\rangle$ being the field vacuum state. Recalling that $N$ is even, Hamiltonian $\hat H_{\rm hop}$ evidently enjoys a mirror symmetry with respect to its middle point, i.e., it is invariant under the
transformation $\hat P| x\rangle\ug | N\meno x\piu1 \rangle$, where $\hat P$ is the parity operator. Thereby, $\hat H_{\rm hop}$ can be block-diagonalized, each block corresponding to a subspace of a given parity (even or odd).
The even (odd) subspace is $N/2$-dimensional and spanned by the basis $\{\left|  x \right>_+\}$ ($\{\left|  x \right>_-\}$) with $|x\rangle_\pm\ug (\left| x \right> \!\pm\! \left|  N\meno x\piu 1 \right>)/ \sqrt{2}$, where $x$ runs from 1 to $N/2$.
For now, we add the requirement that the number of cavities is such that $N/2$ must be odd, which is equivalent
to demand that $N$ -- besides being even -- is not an integer multiple of 4 (for our purposes, this is only a mild restriction).

It is straightforward to check that the parity subspaces introduced above
yield an
effective representation of Hamiltonian (\ref{Hf}) given by
\begin{equation} \label{Hfpm}
\hat H_{\mathrm{hop}}^{(\pm)}\ug - J\!\!\!\sum_{x=1}^{N/2-1}\!\!\![1\!-\!(-1)^{x} \eta] \,(\hat a_{x+1}^{(\pm)\dag}\hat a_{x}^{(\pm)}\piu\mathrm{H.c.})\!\mp\! J_1  \hat a_{N/2}^{(\pm)\dag}\hat a_{N/2}^{(\pm)}
\end{equation}
with $\hat a^{(\pm)\dag}|{\rm vac}\rangle\ug \ket{x}_\pm$
(if $N/2$ is even, an analogous expression holds but replacing $J_1 \!\rightarrow\! J_2 $ on the last term). Note that, unlike in \fig\ref{fig1} where the outermost couplings are equal to $J_1$, 
here the leftmost and rightmost couplings are $J_1 $ and $J_2$, respectively. Thus, Hamiltonian $\hat H_{\mathrm{hop}}^{(\pm)}$ describes an effective array comprising an {\it odd} number of cavities featuring a staggered
pattern of hopping rates {\it and} a defect at the rightmost cavity $x\ug N/2$. This defect consists in a local-frequency shift $\mp J_1$. 

For convenience, let us define $M\ug N/2$ and $\hat V _\pm\ug\mp J_1   \hat a_{M}^{(\rm \pm)\dagger}\hat a_{M}^{(\pm)}$, where the latter describes the defect term in 
Eq. \ref{Hfpm}. 
We can now tackle the problem perturbatively by interpreting $\hat V _\pm$ as a perturbation on a defect-free staggered CCA consisting of an odd number of cavities, a model which can be exactly solved in the single-excitation subspace \cite{ciccarello11}.

\subsection{Diagonalization of $\hat H_{\mathrm{hop}}^{(\pm)}$ for $\hat V_\pm\ug0$}

Based on \rref\cite{ciccarello11}, for $\hat V_\pm\ug0$ (no defect) the spectrum of $\hat H_{\mathrm{hop}}^{(\pm)}$ comprises a pair of bands (separated by a gap $\Delta\omega$) alongside a discrete frequency $\omega_b\ug 0$ falling on the middle of the gap.
The latter corresponds to a bound eigenstate $|\alpha_b \rangle$, which is localized in the vicinity of only one of the array edges (which of the two depends on the sign of $\eta$). This reads 
\begin{equation} \label{boundmode}
\vert \alpha_b   \rangle = \mathcal{C} \sum_{x=1}^{\frac{M+1}{2}} {\mathcal D }^{x-1} \vert 2x-1 \rangle _{\pm}\, 
\end{equation}
with 
\begin{equation} \label{c-tau}
\mathcal D \ug \frac{J_1  }{J_2 } \ug\frac{1\piu\eta}{1\meno\eta}\,,\,\,\,\,\,\,\mathcal{C}\ug \frac{2}{\eta \!- \!1}\sqrt{\frac{\eta}{\mathcal D ^{M\!+\!1}\!-\!1}}\,,\,\,
\end{equation}
where  $\mathcal D $ can be interpreted as the distortion ratio. 
Note that the spatial amplitude of the bound mode, $_{\pm}\langle x | \alpha_b \rangle$,
decays exponentially as $x$ moves away from the weakly-coupled edge. Also,
$_{\pm}\langle x | \alpha_b \rangle\ug0$ for even $\ket{x}_\pm$.
 
All the remaining eigenvalues, instead, are given by
$\omega_{k\mu} \ug - \mu E_{k}$ with $\mu \ug \pm$ (band index) and
\begin{equation}
E_{k} = 2 J \sqrt{ \mathrm{cos}^{2}\frac{k}{2}+\eta^{2}\mathrm{sin}^{2}\frac{k}{2} }\,,
\end{equation}
where $k \ug 2 \pi j / (M\piu1)$ for $j \ug 1,2,\cdot\!\cdot\!\cdot ,(M\meno1)/2$. These describe a pair of energy bands separated by a band gap 
$\Delta \omega \leq 4J$, with the identity holding only when $|\eta| = 1$
The eigenstates corresponding to $\omega_{k\mu}$ are worked out as \cite{ciccarello11}
\begin{align}
\vert \alpha_{k \mu} \rangle & =  \sqrt{\frac{2}{M+1}} \left(  \sum_{x=1}^{\frac{M-1}{2}} 
\mathrm{sin}(kx)\vert 2x \rangle _{\pm} \right. \nonumber \\
 & \quad + \left. \mu  \sum_{x=1}^{\frac{M+1}{2}} \mathrm{sin}(kx + \vartheta _{k})
\vert 2x -1 \rangle _{\pm} \right), 
\end{align}
where the phase $\vartheta _{k}$ is defined by the identity $e^{i\vartheta_{k}} \ug{J(1\meno\eta)}(e^{-ik}\meno\mathcal D )/{E_{k}}$.

\subsection{Peturbative diagonalization of $\hat H_{\mathrm{hop}}^{(\pm)}$}\label{perturb}

Let us now tackle the full problem of diagonalizing
$\hat H_{\mathrm{hop}}^{\pm}$ (taking the defect into account).  
If $J_1  \!\ll\!J_2 $, meaning that the end cavities are 
weakly coupled to the bulk (see \fig\ref{fig1}), $\hat V_\pm$
can be treated as a
small perturbation. 
Applying standard first-order perturbation theory, 
the bound-mode frequency $\omega_b\ug0$ is then straightforwardly corrected as
\begin{eqnarray} \label{boundfreq}
\omega_{b_{\pm}}&\!\simeq\!& \omega_{b} \!\mp\! 
J_1  \langle \alpha_b   \vert \hat a_{M}^{(\rm \pm)\dagger}\hat a_{M}^{(\pm)} \vert \alpha_b   \rangle\ug\mp\!
\dfrac{ 4J \eta\mathcal D ^{M}}{(\eta -1)(\mathcal D ^{M+1}\meno 1)}\,,\,\,\,\,\,\,\,\, 
\end{eqnarray}
where terms $\sim\!O(J_1^2  )$ have been neglected. The perturbation thereby splits $\omega_b$ into two discrete frequencies
separated by the energy gap
\begin{eqnarray} \label{splitting}
\delta\omega=\omega_{b_-}-\omega_{b+}\ug
\dfrac{ 8J \eta}{\eta -1}\,\,\,\frac{\left(\frac{1+\eta}{ 1-\eta}\right)\!^{N/2}}{\left(\frac{1+\eta}{ 1-\eta}\right)\!^{N/2+1}\meno 1}\,,\,\,\,\,\,\,\,\, 
\end{eqnarray}
where we used \eqs (\ref{c-tau}) and (\ref{boundfreq}) and replaced $M\ug N/2$.

\begin{figure}[t!] 
\includegraphics[width=0.48\textwidth]{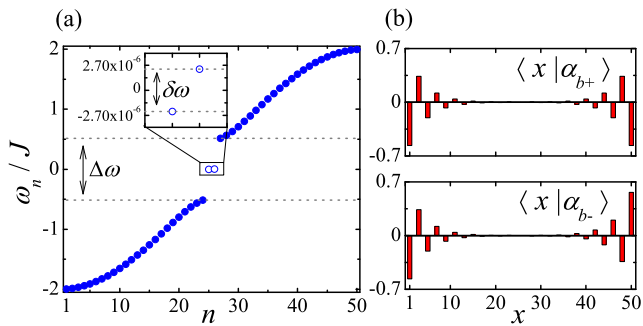}
\caption{\label{fig2} (Color online) (a) Single-excitation spectrum of Hamiltonian (\ref{Hf}) (in units of $J$). $\Delta \omega$ 
is the gap between the pair of bands corresponding to unbound states, while $\delta \omega \ug\omega_{b -} \meno  \omega_{b +}$ 
(displayed in the inset) is the energy gap between the localized bound states [\cf\eq(\ref{splitting})].
(b) Spatial profile of $\ket{\alpha_{b\pm}}$. The plots were obtained by exact numerical diagonalization of \eq(\ref{Hf}) for 
$\eta \ug -0.25$ and $N\ug50$ 
(comparison with perturbation theory \eqs(\ref{boundfreq}) and (\ref{newbound}) is found to be excellent).
}
\end{figure}

The corresponding eigenstates are evaluated as
\begin{align} \label{newbound}
\ket{\alpha_{b_\pm}} & \simeq \vert \alpha_b   \rangle \mp 
 J_1   \sum_{k,\mu} \frac{\langle \alpha_{k\mu} \vert \hat a_{M}^{\dagger}\hat a_{M} \vert \alpha_b  \rangle }{\omega_{b}-\omega_{k\mu}} \vert \alpha_{k\mu} \rangle
\nonumber \\
& = \vert \alpha_b   \rangle  \mp  
4J \mathcal{C} \left(\dfrac{\eta+1}{M+1}\right)\mathcal D ^{\frac{M-1}{2}} \sum_{k} \sum_{x=1}^{\frac{M-1}{2}} \dfrac{\mathrm{sin}(kx)}{E_{k}} \nonumber \\
& \quad \times \, \mathrm{sin}\left[ \left(\dfrac{M\piu1}{2} \right) k\piu\vartheta_{k} \right]
\vert  2x \rangle _{\pm} .  
\end{align}
The unbound states of $\hat H_{\mathrm{hop}}^{(\pm)}$
can be easily obtained as well though they yield extensive expressions
which we do not report here for the sake of brevity.
%
In \fig \ref{fig2}, we consider the paradigmatic instance $\eta\ug-0.25$ and $N\ug 50$, and display the energy spectrum of the full Hamiltonian (\ref{Hf}) alongside the spatial profile of the bound states (\ref{newbound}) on the actual array (i.e., in the basis $\{|x\rangle\}$). We see that the two localized bound states are well-isolated 
from
the unbound modes (the latter corresponding to the pair of bands). They exhibit an energy splitting $\delta \omega$ that, although negligible compared to the band gap $\Delta\omega$, is 
non-zero. Moreover, each bound state is strongly localized in the vicinity of the array edges (i.e., cavities $x\ug1$ and $x\ug N$), a property which 
from now on we refer to as {\it bi-localization}. Those features are
key sources for performing QST, as we discuss next.

\section{Quantum-state transfer: review}\label{review}

QST protocols are typically formulated in one-dimensional XX-type spin chains, which
can be described in terms of ladder spin operators 
yielding a Hamiltonian of the general form
\begin{equation}\label{H-chain}
\hat H_{\rm ch}=\sum_{x=1}^N B_x \hat\sigma_{x}^{+}\hat\sigma_{x}^{-}{+}\sum_{x=1}^{N-1}J_{x}\left( \hat \sigma_{x+1}^{+}  \hat \sigma_{x}^{-}\piu{\rm H.c.}\right)\,,
\end{equation}
where $B_x$ is a local effective magnetic field and $\hat\sigma_{x}^{+} {=}[\hat \sigma_x^{-}]^\dag\ug \ket{1}_x\!\bra{0}$
with $\{|0\rangle_x$, $|1\rangle_x\}$ being a single-spin orthonormal basis.
Note that Hamiltonian (\ref{H-chain}) conserves the total number of excitations, i.e., $[\sum_x \hat \sigma^+_{x}\hat \sigma^-_{x},\hat H_{\rm ch}]\ug0$. 
In the single-excitation
subspace, the Hamiltonian reduces to a tridiagonal matrix describing a standard hopping model.

\subsection{Basics of QST}\label{protocol}

In the usual QST scheme \cite{bose03}, the sender
prepares an arbitrary qubit state $\ket{\phi}_1\ug c_0\ket{0}_1\piu c_{1}\ket{1}_1$ at the first site and
sets the rest of the chain to 
$\ket{0}_2 \cdots\ket{0}_N $.
The initial state of the whole chain thus reads $|\Psi(0)\rangle\ug\ket{\phi}_1 \ket{0}_2 \cdots\ket{0}_N$.
The system then evolves according to its Hamiltonian $\hat H_{\rm ch}$ so that at time $t$ its state is given by $|\Psi(t)\rangle\ug \hat U(t)|\Psi(0)\rangle$ with $\hat U(t)\ug e^{-i \hat H_{\rm ch} t}$.
The goal is to exploit such natural dynamics for transferring the initial sender's state 
$|\phi\rangle$ to the $N$th spin (receiver) in a given time $\tau $, meaning that
$|\Psi(\tau )\rangle\ug \ket{0}_1 \cdots\ket{0}_{N-1}\ket{\phi}_{N}$.
The received (generally mixed) state is evaluated by  
tracing out the remaining spins, i.e., $\rho_N(\tau)\ug\mathrm{Tr}_{1,\ldots, N-1}\ket{\Psi(\tau)}\bra{\Psi(\tau)}$. One thus aims at making the QST fidelity 
$F_\phi(\tau )\ug\bra{\phi}\rho_N(\tau )\ket{\phi}$
as large as possible (the fidelity $F_\phi$ measures how close is the receiver's state to $|\phi\rangle$). 

The fidelity introduced above depends on the specific input $|\phi\rangle$. 
In order to end up with a state-independent figure of merit for QST,
one needs to average $F_\phi$ over all possible input states on
the Bloch sphere ($|c_{0}|^2+|c_{1}|^2=1$). 
For Hamiltonians of the form (\ref{H-chain}), which conserves
the total number of excitations, and given that $|\Psi(0)\rangle$
is restricted to evolve
in the zero- and one-excitation subspaces, the former being unaffected by $U(t)$, 
the {\it average fidelity}
is simply given by \cite{bose03}
\begin{equation}\label{E.avFid}
\mathcal{F}(t )=\frac{1}{2}+\frac{|f (t )|}{3}+\frac{|f (t )|^2}{6}~\,,
\end{equation}
where 
\begin{equation}\label{trans}
f(t)\ug\langle N|e^{-i \hat H_{\rm ch}t}\ket{1}
\end{equation}
is the excitation transition amplitude from the first to the last spin.
(we used the compact notation $\ket{x}\equiv \hat \sigma_{x}^{+}\ket{0}_1 \cdots\ket{0}_N$). 
Note that $|f (\tau)|\ug1$ entails $\mathcal F (\tau)\ug1$ (perfect QST). 
Also, the average fidelity is a monotonic function of the transition amplitude and hence
the QST performance can be evaluated by just tracking down 
the excitation transport across the array. 

When the state to be transferred is encoded in more than two levels (a qutrit for instance) and/or the chain is not properly initialized (thus containing unwanted excitations),
the average fidelity is not expressed by Eq. (\ref{E.avFid}), even though it still depends on the involved transition amplitudes \cite{Latmiraletal15,lorenzo15}. 

\subsection{Rabi-like QST}\label{rabi}

In the single-excitation sector, the spectral decomposition of Hamiltonian (\ref{H-chain}) reads
$\hat H_{\mathrm{ch}}\ug\sum_{j=1}^N \omega_j  |\upsilon _j \rangle\!\langle \upsilon _j |$, where $\omega_j$ is the $j $th energy eigenvalue with corresponding eigenstate 
$\ket{\upsilon _j}\ug\sum_{j=1}^N \upsilon _{jx} \ket{x}$. In this representation, the transition amplitude discussed above is given by
\begin{equation}\label{E.TranAmp}
f (t)\ug\sum_{j=1}^N e^{-i \omega_j  t} \upsilon ^*_{j N} \upsilon _{j 1}\ug\sum_{j =1}^N e^{-i \omega_j  t} \langle \upsilon _j |\hat \sigma_{1}^{+}\hat \sigma^-_{N}|\upsilon _j \rangle.\,\!\!\!\!\!
\end{equation}
The last identity shows that each eigenstate contributes to Eq.~(\ref{E.TranAmp}) through the quantity $\langle \upsilon _j |\hat \sigma_{1}^{+}\hat \sigma^-_{N}|\upsilon _j \rangle$, evolving in time at rate $\omega_j$. In the remainder of this paper, we will refer to 
it as the 
{\it end-to-end amplitude}.

Various high-quality QST schemes \cite{plastinaPRL2007, linneweberetalIJQI2012,lorenzo13,wojcik05,huo08} 
rely on the situation where the edge states $\ket{1}$ and $\ket{N}$ have a strong overlap with only two stationary states, say those indexed by $j\ug1,2$ (bi-localization). In this case, Eq.~(\ref{E.TranAmp}) can be approximated as  
\begin{equation}\label{f2}
f (t)\simeq  e^{-i \frac{\delta\omega\, t}{2}}\langle \upsilon _{1 } |\hat \sigma_{1}^{+}\hat \sigma^-_{N}|\upsilon _{1 }\rangle\piu e^{i\frac{ \delta\omega\, t}{2}}\langle \upsilon _{2 } |\hat \sigma_{1}^{+}\hat \sigma^-_{N}|\upsilon _{2 } \rangle
\end{equation}
with $\delta \omega\ug\omega_{1 }\meno\omega_{2 }$ (we assumed $\omega_{1 }>\omega_{2 })$.
This entails a Rabi-like dynamics that occurs with a characteristic Rabi frequency given by $\delta\omega$. Accordingly, $\tau\sim\delta\omega^{-1}$ showing that the order of magnitude of the transmission time is set by the energy gap between the two bi-localized eigenstates.

The above bi-localization effect is usually achieved by introducing perturbation terms in the Hamiltonian that decouple the outermost spins from the bulk. This can be realized through: (i) application of strong local magnetic fields on the edge spins~\cite{plastinaPRL2007, linneweberetalIJQI2012}, or (ii) on their nearest-neighbours~\cite{lorenzo13}, and (iii) engineering of weak couplings between the edge spins and bulk~\cite{wojcik05,huo08}. While all these models share that a pair of Hamiltonian eigenstates exhibit strong bi-localizaton on the edge sites, the typical energy gap between such two states -- and accordingly the transmission time -- depend on the considered model. 
Calling $\xi$ the model-dependent perturbation parameter (such as the local magnetic field strength),
in (i) the time scales with $N$ as $\tau\sim \xi^N$, resulting in a QST time that exponentially increases with the array length, whereas in (ii) and (iii) the time scales as $O(\xi^2)$ and $O(\xi^{-2})$, respectively. 
All those typical transfer times are in general relatively long and may easily exceed the system's coherence time scale. 
Therefore, it is of great importance to design protocols demanding 
shorter transfer times. 

\section{QST in atom-free staggered CCAs}\label{atomfree}

Comparing \eqs(\ref{Hf}) and (\ref{H-chain}), it should be evident that 
within the single-excitation
subspace, one can regard the spin chain as an atom-free CCA. 
Indeed, in such a case the mapping is straightforward and reads
$\hat\sigma_{x}^{+}\rightarrow \hat a_{x}^{\dagger}$, 
$\hat \sigma_{x}^-\rightarrow \hat a_{x}$.
Likewise, the QST protocol previously discussed in Section \ref{protocol} 
now takes place
in the zero- and one-photon sectors
$\{|{\rm vac}\rangle,\,|x\rangle\}$.
Until Section \ref{modularization} we will thus 
address QST along a staggered CCA with no atoms. This will introduce one of our main results of Section~\ref{modularization}, where we show that the staggered CCA QST time can be significantly reduced by adding {\textit{modularization}} on top of the staggered scheme. On the one hand, this analysis provides
the necessary basis for QST on CCAs coupled to atoms, which we will investigate starting from Section \ref{sec3}. On the other hand, it has its own 
relevance since our findings are independent of the CCA-based implementation, hence they apply to any spin chain with an analogous
pattern of couplings.

\begin{figure}[t!] 
\includegraphics[width=0.45\textwidth]{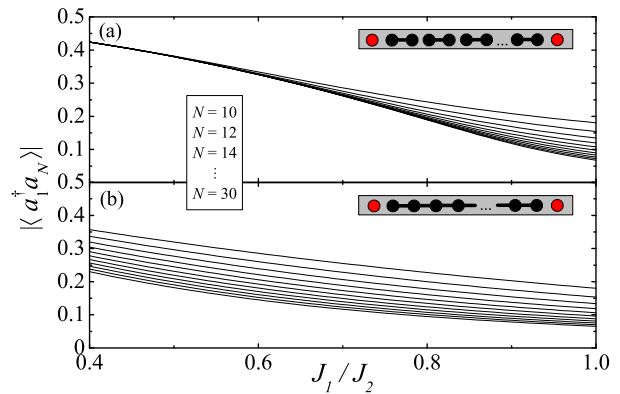}
\caption{\label{fig3} (Color online) End-to-end amplitude $|\langle \alpha_{b\pm} |\hat a_1^\dag \hat a_N |\alpha_{b\pm}\rangle|$ vs.~$J_1/J_2 $ for several values of $N$ (in increasing order from top to bottom) in the case of a staggered array (a) described by Hamiltonian (\ref{Hf}) and a uniform bulk (b) described by Hamiltonian (\ref{Hf-bulk}). Note that $J_1/J_2$ decreases from right to left.  Each plot is obtained from an exact numerical diagonalization of the Hamiltonian.} 
\end{figure}

In the light of Sections \ref{sec2} and \ref{review}, 
the atom-free staggered array 
is suitable for implementing QST based on bi-localization (see Section \ref{rabi}) in the regime $J_1 \ll J_2$. 
To see this, consider first the limiting case
$J_1\ug0$, i.e., $\eta\ug-1$. In this limit (dimerization), the array reduces to a pair of isolated cavities at the outermost sites and a bulk of uncoupled dimers [see the small sketch on top of \fig\ref{fig3}(a)]. 
The pair of
bound states [\cf\eq(\ref{newbound})] then reduce to the doublet $\ket{\alpha_{b\pm}}\ug( |1\rangle\pm |N\rangle)/\sqrt{2}$ with $\omega_{b_{\pm}}\ug0$, these being evidently the only stationary states with non-zero amplitude
at the array ends. 
This would turn \eq(\ref{f2}) into an exact identity with $\{\ket{\alpha_{b\pm}}\}$ embodying the pair $\{\ket{\upsilon_1},\ket{\upsilon_2}\}$. 
Yet, due to $\omega_{b_{\pm}}\ug0$, 
the transmission time $\tau$ would be infinite since $\delta\omega\ug0$. To make this finite, we thus need to work in the regime $J_1\!\ll\!J_2$, which justifies our perturbative approach in Section \ref{perturb}.
 
Next, with the help of \eqs(\ref{boundmode}) and (\ref{newbound}), we note that the end-to-end amplitudes entering \eq(\ref{f2}) fulfill
\begin{equation}\label{ete}
\langle \alpha_{b\pm} |\hat a_1^\dag \hat a_N|\alpha_{b\pm}\rangle \ug \pm\tfrac{\mathcal{C}^2}{2}+O(J_1^2)\,.
\end{equation}
Thereby, the transition amplitude's modulus reads 
\begin{equation}\label{f2-stagg}
|f (t)| = 2\left|\langle\hat a_1^\dag \hat a_N\rangle\,\mathrm{sin}\left(\dfrac{\delta \omega }{2}t\right)\right| ,
\end{equation}
where $\langle\hat a_1^\dag \hat a_N\rangle$ is a short notation for the end-to-end amplitude. 
At times $t\ug 2n\pi /\delta\omega$ with $n$ being an odd integer, 
Eq. (\ref{f2-stagg}) reaches the value $ 2|\langle\hat a_1^\dag \hat a_N\rangle|$. Hence, ideally, if the absolute value of the end-to-end amplitude
equals 1/2, perfect QST is attained with transmission time $\tau\ug2\pi/\delta\omega$.

In \fig\ref{fig3}(a), based on exact numerical diagonalization of \eq(\ref{Hf}), we explore how the end-to-end amplitude is affected by the array size $N$ and $J_1/J_2$.
For a set ratio $J_1/J_2$, the amplitude decreases with $N$, eventually saturating to an asymptotic value. For $J_1/J_2\ug1$ (uniform hopping rates) the asymptotic value is well below 1/2 but tends to it as $J_1/J_2$ approaches zero. At the same time, remarkably, the rapidity at which $|\langle \hat a_1^\dag \hat a_N\rangle|$ saturates to such asymptotic value as a function of $N$ grows up in a way that, for $J_1/J_2$ small enough, the amplitude becomes in fact  $N$-independent. This agrees with \eq(\ref{ete}) [see also \eq(\ref{c-tau})].
%

In other words, for a very distorted array, the bi-localization effect required for high-fidelity QST is about insensitive to the system size. 
This property is related to what is known as true long-distance entanglement exhibited by the ground state
of staggered spin chains \cite{venuti07}, as
opposed to quasi-long-distance entanglement featuring quantum correlations that decrease with $N$.
The latter occurs, for instance, in spin chains comprising a uniform bulk \cite{wojcik05, venuti07}. The fundamentally different nature
of those two situations reflects in the scaling properties of QST fidelity as well. To show this, consider
a CCA where -- unlike the staggered array -- the bulk cavities are coupled uniformly with rate $J_2$ [see sketch on top of \fig\ref{fig3}(b)]. The Hamiltonian of such an array thus reads
\begin{equation}\label{Hf-bulk}
\hat H_{\mathrm{hop}}' \ug -J_1  (\hat a_{2}^{\dagger}\hat a_{1}\piu \hat a_{N}^{\dagger}\hat a_{N-1})\meno J_2 \!\sum_{x=2}^{N-2}\hat a_{x+1}^{\dagger}\hat a_{x} + \mathrm{H.c.}
\end{equation}
Since the outermost sites are weakly coupled to the bulk,
a pair of bi-localized eigenstates is formed in this case too \cite{wojcik05}.
In Fig. \ref{fig3}(b), we plot the corresponding end-to-end amplitude as a function of $J_1/J_2$ and $N$. The differences with respect to the
staggered-CCA case are quite striking. While for $J_1/J_2\ug1$ (fully uniform array) 
both
models coincide, the end-to-end amplitude in the uniform-bulk case 
decreases with $N$ at variance with the stable behaviour found in the staggered model,
taking moreover lower values compared to the latter. 
This shows
some of the
attractive features of staggered arrays in terms of QST fidelity.

\section{Modularized array}\label{modularization}

The advantages highlighted in the previous section, however, come
with a price in terms of
the transmission time $\tau$ required for carrying out QST. 
Recalling that $\tau \sim \delta\omega^{-1}$, 
\eq(\ref{splitting}) indeed shows that, in the regime $J_1\!\ll\! J_2$ (i.e., $\eta\!\simeq\!-1$), the bound-state gap $\delta\omega$ 
exponentially decays with the size $N$. As a consequence, $\tau$ exponentially grows up with $N$. 
One thus wonders whether, for a given size, the staggered array can be modified so as to increase the
gap while maintaining the bi-localization strength of $\ket{\alpha_{b\pm}}$ (necessary to attain high fidelity). In this
section, we show that this can be achieved by {\it modularizing} the staggered CCA.

The setup we put forward is inspired by the concept of modular entanglement
introduced in \rref\cite{gualdi11}. Let us consider then a set of $m$
identical staggered arrays, having $N$ sites each, so that the total
number of sites is $L \ug mN$. 
Nearest-neighbour cavities of adjacent modules are coupled with 
hopping rate $J_{\rm mod}$, hence the
total Hamiltonian reads
\begin{equation}\label{Hmod}
\hat H_{\mathrm{mod}} \ug \sum_{j=1}^{m}\hat H_{\mathrm{hop}}^{(j)}-J_{\mathrm{mod}}\!
\sum_{j=1}^{m-1}(\hat a_{jN+1}^{\dagger} \hat a_{jN}+\mathrm{H.c.})\,,
\end{equation} 
where the free module Hamiltonian $\hat H_{\mathrm{hop}}^{(j)}$ is the same as \eq(\ref{Hf}) [the sum being now over $x\ug(j\!-\!1)N\piu1, jN-1$].

For $J_{\rm mod}\ug J_2$, the whole setup reduces to a standard staggered array comprising $L$ cavities. In contrast, in the limit $J_{\rm mod}\ug0$ (no inter-modular couplings), 
the energy spectrum and associated eigenstates of $\hat H_{\rm mod}$ are the same as those of a single $N$-long module analyzed in Section \ref{sec2}, 
but becoming $m$-fold degenerate.  For intermediate values $0<J_{\rm mod}<J_2$, such degeneracy is removed 
resulting in a manifold of 2$m$ non-degenerate bound states. Among these, let us call $\delta \omega_{m,N}$ the energy gap between the pair of most internal ones
and $|\langle\hat a_1^\dag \hat a_L\rangle|$ the absolute value of their end-to-end amplitude. Then, for $J_{\rm mod}\ug J_2$, $\delta \omega_{m,N}$ and $|\langle\hat a_1^\dag \hat a_L\rangle|$ 
are respectively the same as $\delta\omega$ and the corresponding end-to-end amplitude of a staggered array of size $L$ [see \eq(\ref{splitting}) and \fig\ref{fig3}(a)]. In the opposite limit $J_{\rm mod}\ug 0$, 
$\delta \omega_{m,N}$ is {\it larger}, since now it coincides with the bound-state gap of a staggered array of size $N\!<\!L$, while $|\langle\hat a_1^\dag \hat a_L\rangle|\ug0$ because the modules 
are now uncoupled.

\begin{figure}[t!] 
\includegraphics[width=0.48\textwidth]{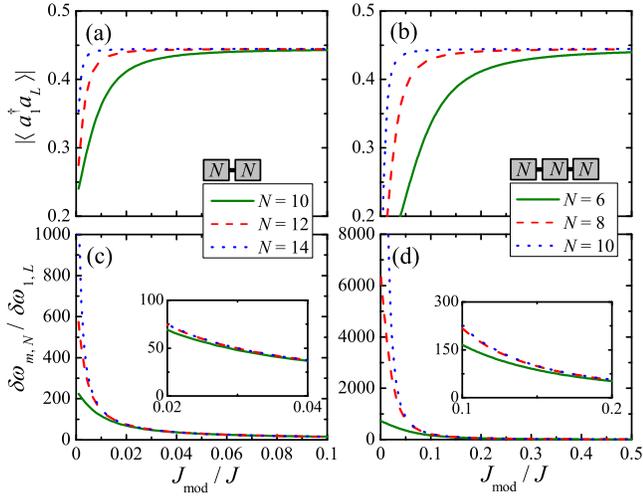}
\caption{\label{fig4} (Color online) Full-array end-to-end amplitude $|\langle\hat a_1^\dag \hat a_L\rangle|$ [(a) and (b)]
and energy-gap gain $\delta \omega_{m,N}/\delta\omega_{1,L}$ [(c) and (d)] against $J_{\mathrm{mod}}/J$ for different values of $N$
in the case of a modularized staggered CCA. Left-column plots [(a) and (c)] refer to a two-module
array ($m\ug2$), while right-column ones [(b) and (d)] correspond to a three-module array ($m\ug3$). 
For each setup, we have set the intra-module distortion to $\eta \ug -0.5$ (about $J_{1}/J_{2} = 0.33$).}
\end{figure}

To investigate the dependence of $\delta \omega_{m,N}$ and $|\langle\hat a_1^\dag \hat a_L\rangle|$ on $J_{\rm mod}$, in \fig\ref{fig4} we consider
the cases of a two- and three-module array ($\delta \omega_{m,N}$ is plotted in units of $\delta \omega_{1,L}$, namely its value at $J_{\rm mod}\ug J_2$). As $J_{\rm mod}$ grows from zero, both the gap
and the end-to-end amplitude monotonically tend to their respective values for $J_{\rm mod}\ug J_2$ (i.e., the case discussed above). 
Remarkably,
the end-to-end amplitude in particular exhibits quite a fast saturation [see \figs\ref{fig4}(a) and (b)]. Instead, $\delta\omega_{m,N}$ undergoes a more regular growth.
This means that, starting from $J_{\rm mod}\ug J_2$ ($L$-size staggered array) one can 
decrease $J_{\rm mod}$ by a significant amount -- thus modularizing the CCA -- and keep the end-to-end amplitude about unchanged but
amplifying the energy gap substantially. 
For instance [see \figs\ref{fig4}(a) and (c)], in the two-module ($m\ug2$) case for $N\ug14$ when $J_{\rm mod}\simeq0.01 J$ the end-to-end amplitude is unchanged 
for all practical purposes while the energy gap is over a hundred times larger, resulting in the same QST fidelity but with a transfer time about two orders of magnitude lower. 
This can be further improved
by increasing the number of modules, for fixed overall array length $L$ since this results in modules of shorter length.

We also note from \fig\ref{fig4} that the saturation of $|\langle\hat a_1^\dag \hat a_L\rangle|$  occurs for lower values of $J_{\rm mod}$ as $N$ grows. Hence, 
lower values of $J_{\rm mod}$ are required for establishing bi-localization. This can be attributed to the fact that the gap $\delta \omega$ of each (isolated) staggered module, coinciding with $\delta\omega_{m,N}$ for $J_{\rm mod}\ug0$, decreases with $N$. 
From a perturbative perspective, the effect of switching on an inter-modular coupling will be significant when $J_{\rm mod}$ becomes comparable with $\delta \omega$ which, however, decreases with $N$.

\begin{figure}[t] 
\includegraphics[width=0.45\textwidth]{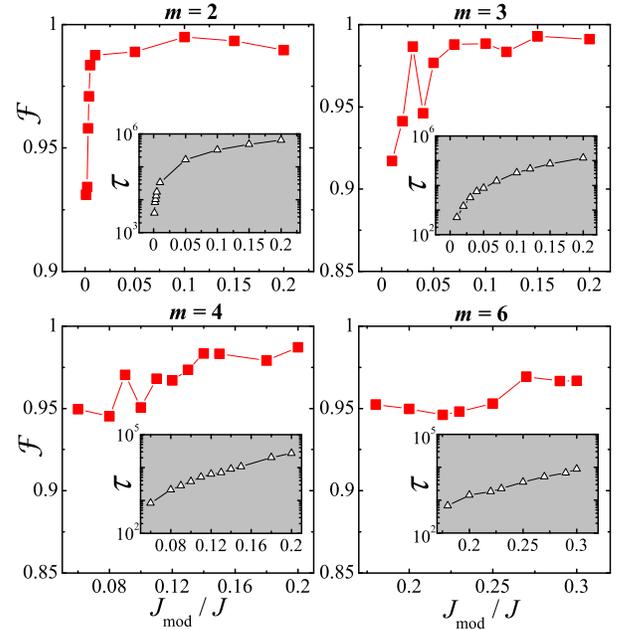}
\caption{\label{fig5} (Color online) Maximum achievable average QST fidelity $\mathcal{F}$ [\cf Eq.~\ref{E.avFid})] after one Rabi-like oscillation period, that is $\tau\ug2\pi/\delta \omega_{m, N}$, against $J_{\rm mod}/J$. We have set $L\ug24$, $J_1\ug 0.3 J$ and $J_2\ug J$ and considered different modularization schemes (each specified by the value of $m$).
In each panel, the inset shows the transfer time $\tau$ (in units of $J^{-1}$) vs.~$J_{\rm mod}/J$ in a log-lin scale. 
For the unmodularized array ($m\ug1$), the maximum fidelity and transfer time are, respectively, $\mathcal{F}\simeq 0.98$ and  $\tau\simeq 3\cdot 10^6 J^{-1}$.}
\end{figure}

To summarize, for a staggered array of a given length, partitioning it into 
several module can result in shorter QST times without significantly affecting the corresponding fidelity. 
In Fig.~\ref{fig5}, we provide further explicit evidence of this phenomenon by considering a CCA of length $L\ug24$ in the case of four different modularizations defined by $m\ug2,3,4,$ and 6. Note, for instance, that 
a six-block modularization leaves the fidelity above $\simeq 95\%$ while the QST time is shortened by three orders of magnitude. 
A significant QST speed-up is nevertheless attainable even for lower $m$.
Note that while the QST time increases polynomially with $J_{\mathrm{mod}}$, the fidelity shows a non-monotonic behavior due to residual contributions from 
other eigenstates to the transition amplitude [see Eq. (\ref{E.TranAmp})].

\begin{figure}[t] 
\includegraphics[width=0.45\textwidth]{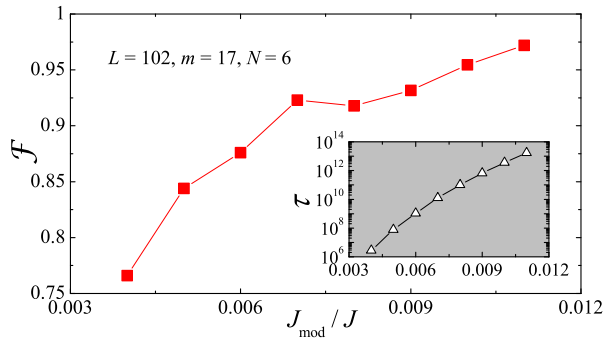}
\caption{\label{figR1} (Color online) Maximum achievable average QST fidelity $\mathcal{F}$ [\cf Eq.~\ref{E.avFid})] after one Rabi-like oscillation period, that is $\tau\ug2\pi/\delta \omega_{m, N}$, against $J_{\rm mod}/J$. We have set $L\ug102$ and $\eta = -0.8$ (about $J_{1}/J_{2} = 0.11$)
using the modularization scheme $m=17$, the length of each module thus being $N=6$.
The inset shows the transfer time $\tau$ (in units of $J^{-1}$) vs.~$J_{\rm mod}/J$ in a log-lin scale. For the corresponding unmodularized CCA, the transfer time is infinite for all pratical purposes.}
\end{figure}

The possibility to reduce QST times over relatively short distances -- say of the order of up to 30 sites as in Fig.~\ref{fig5} -- is relevant itself, e.g., to carry out short-haul communications tasks between quantum processors in a quantum computing architecture. Concerning longer CCAs, a thorough analysis of the scalability of a modularized array is beyond the scopes of the present work and will thus be presented elsewhere \cite{almeida16prep}. However, in order to test the potential of modularized chains to perform QST over longer distances, in Fig.~\ref{figR1} we additionally consider the paradigmatic case of a CCA having $L\!=\!102$ sites. Note that high-quality QST is still achievable within times that, although inevitably longer, are far shorter compared to the unmodularized staggered CCA. 

As mentioned previously, all the above clearly applies not only to atom-free CCAs, 
but spin chains in general (regardless of 
their implementation).
In the following, we will address CCAs coupled to atoms with the goal of putting forward QST schemes in which
both atomic and photonic degrees of freedom are involved.

\section{\label{sec3}CCA with atoms}

We now consider a CCA, where each cavity is additionally coupled to a two-level atom of frequency $\omega_a$, according to the Jaynes-Cummings (JC) interaction Hamiltonian \cite{jaynes63}
\begin{equation}\label{jc}
\hat H_{x}^{(\mathrm{JC})} \ug \omega_c \hat a_{x}^{\dagger}\hat a_{x} \piu \omega_a \hat \sigma_{x}^{+}\hat \sigma_{x}^{-} \piu
g\,(\hat \sigma_{x}^{+}\hat a_{x}\piu\hat\sigma_{x}^{-} \hat a_{x}^{\dagger}),
\end{equation}
where now $\hat \sigma_{x}^{+}\equiv \ket{\mathrm{e}}_x\!\bra{\mathrm{g}}$
with $| \mathrm{g} \rangle$ ($| \mathrm{e} \rangle$) denoting the atomic ground (excited) state, and 
$g$ is the atom-field coupling strength.
In the following, we again set $\omega_c\ug0$ for simplicity.
For a staggered pattern of hopping rates (see \fig\ref{fig1}), the total Hamiltonian reads
\begin{equation} \label{Htot}
\hat H=\hat H_{\mathrm{hop}}+\sum_{x=1}^{N}\hat H_{x}^{\mathrm{(JC)}} ,
\end{equation}
where the hopping Hamiltonian is the same as in Eq. (\ref{Hf}).
%
Hereafter, we adopt the short notation 
$|1_x\rangle\equiv \hat a_{x}^{\dagger}\ket{\mathrm{vac}}\ket{\mathrm{g}}_{1}\cdots\ket{\mathrm{g}}_{N}$ 
and $|e_x\rangle \equiv \hat \sigma_{x}^{+}\ket{\mathrm{vac}}\ket{\mathrm{g}}_{1}\cdots\ket{\mathrm{g}}_{N}$, where 
the former is the state where a single photon lies at the $x$th cavity with
all the atoms unexcited, while in the latter state only the $x$th atom is excited (with the field and all of the remaining atoms unexcited).
The single-excitation sector of the joint Hilbert space is 2$N$-dimensional and spanned by the basis
$\lbrace \ket{1_x}, \ket{e_x}  \rbrace$.  

Moreover, let us denote $\{\ket{\alpha_n}\}$ as the set of $N$ eigenstates of the {\it free} field Hamiltonian $\hat H_{\rm hop}$, i.e., $\hat H_{\rm hop}\ket{\alpha_n}\ug\omega_n\ket{\alpha_n}$, each
having the form $\ket{\alpha_n}\ug\sum_x\alpha_{nx} \left| 1_x \right>$.
These states solely comprise photonic excitations (index $n$ is intended to run over both bound and unbound states). 
Correspondingly, one can define a set of $N$ states $\{\ket{\beta_n}\}$ such that 
$\ket{\beta_n}\ug\sum_x\alpha_{nx} \left| e_x \right>$, hence featuring only atomic excitations (excitons). 
By construction, each $\ket{\beta_n}$ has the same spatial profile as $\ket{\alpha_n}$ and can thus be regarded as its 
excitonic analogue. States $\{\ket{\alpha_n}\}$ ($\{\ket{\beta_n}\}$) can be regarded as arising from the normal-mode field (atomic) operators $\{\hat\alpha_n\}$ ($\{{\hat \beta_n}\}$) defined accordingly as $\hat \alpha_n\ug \sum_x \alpha_{nx}\hat a_x$ ($\hat \beta_n\ug \sum_x \alpha_{nx}\hat \sigma_{x}^{-}$).

Note that in \eq (\ref{Htot}) both $g$ and $\omega_a$ are uniform throughout the array. Using this, $\hat H$ can be rearranged as (see Refs. \cite{ogden08,makin09,ciccarello11})
\begin{equation} \label{Hnormal}
\hat H \ug\! \sum_n\left[ \omega_{n} \hat\alpha_n^\dag\hat\alpha_n\piu\omega_a\hat\beta_n^\dag\hat\beta_n
\piu g(\hat\beta_{n}^\dag\hat\alpha_n\piu{\rm H.c.})\right]\,.
\end{equation}
Therefore, within the single-excitation sector, the system behaves as a set of $N$ {\it decoupled} effective JC models, each corresponding to 
a photonic mode of frequency $\omega_n$ coupled to its excitonic counterpart of frequency $\omega_a$ with coupling strength $g$. 
This allows for a straightforward diagonalization of $\hat H$ once the
eigenstates of the free field Hamiltonian $\hat H_{\rm hop}$, $\{\ket{\alpha_n}\}$, are known. 
Using the standard JC-model theory, indeed, the eigenstates are worked out as 
\begin{equation}\label{dressed}
\ket{ \psi_{n}^{(\pm)} } = A_{n}^{(\pm)} \left| \alpha_{n} \right>
+ B_{n}^{(\pm)} \left| \beta_{n} \right>,
\end{equation}
where
\begin{equation}\label{dressed_ab}
A_{n}^{(\pm)} \ug \dfrac{2g}{\sqrt{(\Delta_{n}\! \pm\! \Omega_{n})^2\piu4g^{2}}},\,\,\,B_{n}^{(\pm)}  \ug \dfrac{\Delta_{n} \!\pm\! \Omega_{n}}{\sqrt{(\Delta_{n}\! \pm\! \Omega_{n})^2\piu4g^{2}}},
\end{equation}
with $\Delta_{n} \ug \omega_a\meno \omega_{n}$ and $\Omega_{n} \ug \sqrt{\Delta_{n}^{2} \piu 4g^{2}}$ being the detuning and vacuum Rabi frequency, respectively,
of the $n$th effective JC model. The corresponding energy levels read
\begin{equation}\label{dressed_energy}
\varepsilon_{n}^{(\pm)}=\tfrac{1}{2}\,(\omega_a+\omega_{n} \pm \Omega_{n}).
\end{equation}

\subsection{\label{single}Single-mode resonance}

Out of all the $N$ effective JC dynamics [\cf\eq(\ref{Hnormal})] one can selectively
excite only one of them upon a judicious tuning of the atomic frequency $\omega_a$. 
Now we particularly show how to
trigger only the JC dynamics corresponding to the bound eigenstate $\ket{\alpha_{b+}}$ [\cf\eq(\ref{newbound})].
In the interaction picture, Hamiltonian (\ref{Hnormal}) is turned into (we now highlight explicitly the contributions of the bound and unbound states)
\begin{eqnarray} \label{Hint}
\hat H_{\rm I} (t)\ug g\!\left[ \sum_{j\ug\pm}\! \hat\beta_{bj}^\dag\hat \alpha_{bj}e^{i\Delta_{bj} t}\!\piu\! \sum_{k\mu} \! \hat\beta_{k\mu}^\dag\hat \alpha_{k\mu}e^{i\Delta_{k\mu} t}\!\piu{\rm H.c.}\right]\,\,\,\,\,\,\,\,\,\,
\end{eqnarray}
with $\Delta_{b\pm}\ug\omega_a\meno \omega_{b\pm}$ and $\Delta_{k\mu}\ug\omega_a\meno \omega_{k\mu}$.
By tuning $\omega_a$ on resonance
with $\omega_{b+}$, namely setting $\omega_a\!=\!\omega_{b+}$ the first term becomes time-independent.
If, additionally, $g\!\ll\{\Delta_{k\mu},\Delta_{b-}\}$ all the remaining terms in \eq (\ref{Hint}) are rapidly rotating so that
they effectively do not affect the dynamics and, hence, can be neglected. Returning to the Schr\"{o}dinger picture, 
we thus end up with an effective Hamiltonian of the form 
\begin{equation} \label{Heff1}
\hat H_{\rm eff} \ug\! \sum_{n}\left( \omega_{n} \hat\alpha_n^\dag\hat\alpha_n\piu\omega_a\hat\beta_n^\dag\hat\beta_n\right)
\piu g(\hat\beta_{b+}^\dag\hat\alpha_{b+}\piu{\rm H.c.})\,.
\end{equation}
An analogous conclusion holds if we set the atomic frequency on 
resonance with $\omega_{b-}$.
The dynamics thus consists of a resonant JC-like dynamics involving $|\alpha_{b+}\rangle$
and its excitonic analogue, while all the remaining photonic and atomic modes
evolve freely.
Accordingly, only the pair of dressed states $\ket{\psi_{b+}^{(\pm)}}$ are thus formed [\cf\eq(\ref{dressed})]. Note that, due to the resonance condition $\Delta_{b+}\ug0$, we get
$|A_{b+}^{(\pm)}|\ug |B_{b+}^{(\pm)}|$ [\cf\eq(\ref{dressed_ab})]. Hence, $\ket{\psi_{b+}^{(\pm)}}$ are {\it fully} dressed states featuring maximal atom-photon entanglement.

\subsection{\label{strong}Strong-coupling regime}

Clearly, an implicit requirement for the above regime to hold is that $g\!\ll\!\delta\omega$ (since $\ket{\alpha_{b-}}$ is the nearest state in energy). If not, additional coupling terms between field modes and 
the respective excitonic analogues would appear in \eq(\ref{Heff1}). Consider, in particular, the strong-coupling regime \cite{makin09,almeida13} such that $g$ is far larger than the 
entire range of the field frequencies ($\omega_a\ug0$ for simplicity). Then, none of the coupling terms in \eq(\ref{Hnormal}) can be neglected in a way that each corresponding JC
dynamics is activated. Also, due to the negligible detunings, {\it all} the pairs of states in Eq. (\ref{dressed}) are formed, each reading $\ket{ \psi_{n}^{(\pm)} }\!\simeq\!  (\left| \alpha_{n} \right> \pm
\left| \beta_{n} \right>)/ \sqrt{2}$, thus embodying fully dressed states. Accordingly, the energy spectrum [\cf\eq(\ref{dressed_energy})] reduces to $\varepsilon_{n}^{(\pm)}\!\simeq{\omega_{n}}/{2}\pm g$ (since $\Omega_n\!\simeq\!2g$). Thereby, in this regime two independent polaritonic bands are formed, each corresponding to even (odd) dressed states $\ket{\psi_n^{(+)}}$ $(\ket{\psi_n^{(-)}})$. In either of these, 
the dynamics thus reduces to a single polariton subjected to an effective Hamiltonian that is analogous to the free field hopping Hamiltonian (\ref{Hf}) [or (\ref{Hmod}) in the case of modularization] but
with all the hopping rates rescaled by a 1/2 factor. If the CCA is prepared in a state such as $(\ket{e_1}\pm\ket{1_1})/\sqrt{2}$, then only the corresponding band will be excited and the dynamics
will be the same as that analyzed in previous sections (with each single-photon state $\ket{x}$ now replaced by the single-cavity polariton state $(\ket{e_x}\!\pm\!\ket{1_x})/\sqrt{2}$).

\section{\label{sec5}Transfer of atomic and polaritonic states}

Depending on the single-mode resonance or strong-coupling regimes discussed in Sections \ref{single} and \ref{strong}, respectively, we now show that one
can carry out transfer of an atomic or polaritonic state.

\subsection{Atomic QST through single-mode resonance}

Setting $\omega_a\ug\omega_{b+}$ and $g\!\ll\!\delta\omega$, the latter being the gap between the bi-localized states $\ket{\alpha_{b\pm}}$, the JCH
Hamiltonian takes the effective form of \eq(\ref{Heff1}). If the parameters entering \eq(\ref{Hf}) [or \eq(\ref{Hmod}) for modularized CCAs] are such that strong bi-localization occurs (see Sections \ref{sec2}, \ref{atomfree}, and \ref{modularization}), then both the excitonic states $\ket{e_1}$ and $\ket{e_N}$ can be decomposed to a good approximation only in terms of $\ket{\beta_{b\pm}}$. This gives $\ket{e_1}\!\simeq\!\sum_{j=\pm}\langle \beta_{bj}|e_1\rangle\ket{\beta_{bj}}$ and, using the parity properties of $\ket{\beta_{b\pm}}$, $\ket{e_N}\!\simeq\!-\langle \beta_{b-}|e_1\rangle\ket{\beta_{b-}}\piu \langle \beta_{b+}|e_1\rangle\ket{\beta_{b+}}$. Expressing
next $\ket{\beta_{b+}}$ in terms of dressed states  [see \eq(\ref{dressed})], we get $\ket{\beta_{b+}}\ug\tfrac{1}{\sqrt{2}}\left(\ket{\psi_{b+}^{(+)}}\meno \ket{\psi_{b+}^{(-)}}\right)$, where $\ket{\psi_{b+}^{(\pm)}}$ has energy
$\omega_0\pm g$. Replacing it into the above decomposition for $\ket{e_1}$ and letting this evolve in time through to the usual time-evolution operator $\hat U(t)$, we get
\begin{eqnarray}
\hat U(t) \ket{e_1}&\!\simeq\!& \langle \beta_{b-}|e_1\rangle\ket{\beta_{b-}}\nonumber\\
&&\piu\tfrac{\langle \beta_{b+}|e_1\rangle}{\sqrt{2}}\left( e^{-igt} \ket{\psi_{b+}^{(+)}}\meno e^{igt}  \ket{\psi_{b+}^{(-)}} \right) 
\end{eqnarray}
up to an irrelevant global phase factor. 
Expressing now again the dressed states in terms of $\ket{\alpha_{b+}}$ and $\ket{\beta_{b+}}$,
\begin{eqnarray}
\hat U(t) \ket{e_1}&\!\simeq\!& \langle \beta_{b-}|e_1\rangle\ket{\beta_{b-}}\nonumber\\
&&\piu\langle \beta_{b+}|e_1\rangle\left[ \cos(gt)\ket{\beta_{b+}}\!-\!i\sin(gt)\ket{\alpha_{b+}} \right] .
\end{eqnarray}
For $gt\ug \pi$ (up to an irrelevant global phase factor), we thus get (see above) $\hat U(t)\ket{e_1}\!\simeq\! \ket{e_N}$.
Noting that, in the light of Section \ref{review}, the state in which the CCA has zero excitations (both photonic and atomic) does not evolve, the
two-level atom constitutes a natural choice for encoding the logical qubit.
Therefore, a QST protocol can be carried out 
between the outermost atoms in a transfer time $\tau\ug \pi/g$. 
%
\begin{figure}[t!] 
\includegraphics[width=0.32\textwidth]{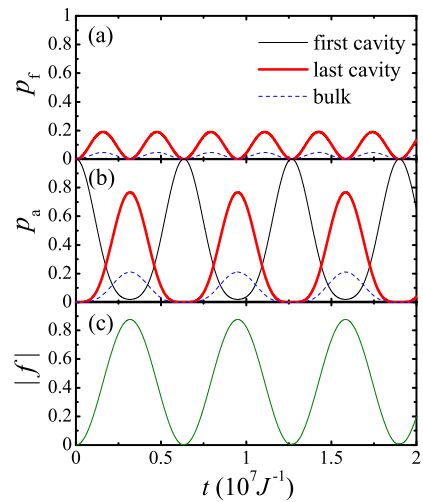}
\caption{\label{fig6} (Color online) 
Time evolution of the photonic (a) and atomic (b) excitation and of the transition amplitude (c) across a 10-cavity 
staggered CCA for an initial state $\ket{\Psi(0)}\ug\ket{e_1}$. In (a) [(b)], we display the probability to find the photonic (atomic) excitation
at the first cavity (thin black), the last one (thick red), and in the bulk sites $2\!\le\!x\!\le\!N-1$ (dashed blue). The plots are obtained from
an exact numerical diagonalization of Eq. (\ref{Htot}) for $\eta\ug-0.5$ and $g\ug10^{-6}J$.}
\end{figure}
Likewise, one can accordingly define a transition amplitude (\cf Section \ref{review}) as $f(t)\ug \langle e_N|\hat U(t)|e_1\rangle$ and evaluate the QST efficiency using
Eq. (\ref{E.avFid}) for the average fidelity. 

In \fig\ref{fig6}, we study in
a paradigmatic instance (such that $\ket{\Psi(0)}\ug \ket{e_1}$) the time evolution of the photonic and atomic excitations alongside 
the transition amplitude just introduced. We denote $p_{{\rm f},x}(t)\ug |\langle 1_x|\Psi(t)\rangle|^2$ and $p_{{\rm a},x}(t)\ug |\langle e_x|\Psi(t)\rangle|^2$ as the
probability to find one photon and one exciton at cavity $x$, respectively. 
As shown in \fig\ref{fig6}, the transfer takes place through
the involvement of the entire CCA, including the bulk (especially in the form of excitons). Note that, while the considered array is only moderately distorted (we take $\eta\ug-0.5$), 
$|f|$ attains a maximum $\simeq 0.9$.   

\subsection{Polariton transmission in the strong-coupling regime}

Note that in the scheme discussed previously, the transfer time $\tau$ is set in fact by the atom-field
coupling strength $g$, which is required to be much smaller than the energy gap between bi-localized modes $\delta\omega$.
As the latter decreases with the array distortion (see Section \ref{sec2}), such a scheme can be demanding for highly-distorted CCAs.
In this scenario, the properties of an atom-free CCA as seen in Sections \ref{sec2}, \ref{atomfree}, and \ref{modularization} can be exploited
to transfer polaritonic states across the array.

In the strong-coupling regime (see Section \ref{strong}), the dynamics
reduces to that of a pair of fully-dressed polaritonic bands. In either of these, a single-cavity polariton
of given parity hops through the array just like a photon propagates
through an atom-free CCA (see Sections \ref{sec2}, \ref{atomfree} and \ref{modularization}) 
apart from a 1/2 factor rescaling of hopping rates (hence half of the propagation
speed). 
Given that the polaritonic bands are uncoupled, the preparation of a polariton
of a given parity in a given cavity, say $\tfrac{1}{\sqrt{2}}(\ket{e_1}\piu\ket{1_1})$, will trigger a dynamics
where solely polaritons of the same parity are involved. Hence, at least in principle, one can encode
a qubit in each cavity in terms of atom-photon logical states 
$\ket{\mathrm{vac}}\ket{\mathrm{g}}_{1}\cdots\ket{\mathrm{g}}_{N}$ and 
$\tfrac{1}{\sqrt{2}}(\ket{e_x}\piu\ket{1_x})$.
Accordingly, in such a framework and in virtue of Section \ref{review}, this leads to a 
transition amplitude defined as $f(t)\ug\tfrac{1}{2} (\bra{e_N}\piu\bra{1_N})\hat U(t)(\ket{e_1}\piu\ket{1_1})$. 
Regardless of the feasibility of such a qubit implementation, $f(t)$ can be used as a 
figure of merit for measuring how reliably a polaritonic state can be transmitted across the CCA in
line with other studies \cite{nohama07, bose07-qst,ogden08,makin09}.

\begin{figure}[t!] 
\includegraphics[width=0.4\textwidth]{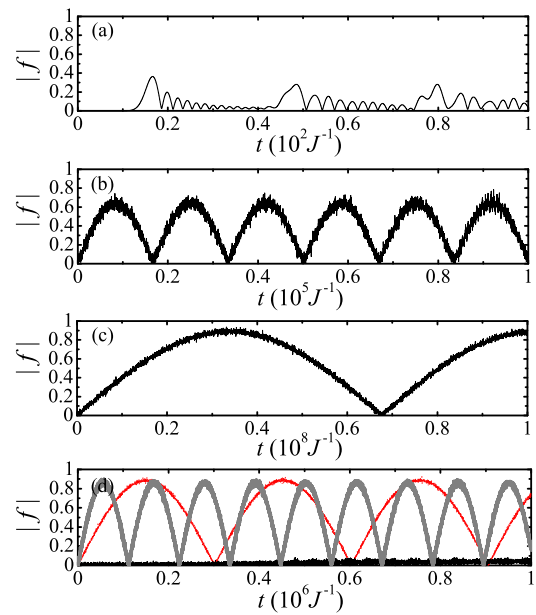}
\caption{\label{fig7}
Time evolution of the transition amplitude for 
an initial symmetric polariton set at the first cavity in
the case of a staggered 30-cavity array for (a)  $\eta \ug 0$, 
(b) $\eta\ug -0.25$, and (c) $\eta \ug -0.5$ (solid black). 
In (d) we show the case of a modularized CCA for $m=3$ with $J_{\mathrm{mod}} = 0.1J$ (dotted red) and $m=5$ with $J_{\mathrm{mod}} = 0.3J$ (thick gray). Note that $J_{\mathrm{mod}}$ was slightly  increased in order to assure the formation of bi-localized states (\cf Section \ref{modularization}).
The intra-modular distortion parameter was fixed to $\eta \ug -0.5$.
We set $g\ug0.01J$ and $\omega_a\ug0$ throughout. Plots are
obtained from an exact diagonalization of Eq. (\ref{Htot}) [with $\hat H_{\mathrm{hop}}$ being replaced with $\hat H_{\mathrm{mod}}$ in (d)].
}
\end{figure}

Interestingly, in order for the polariton transfer to be effective, the requirement that $g$ must be strong enough  
in order to enable the entire set of dressed states to form is not strict. Indeed, the nature 
of QST across
an atom-free CCA investigated in Sections \ref{sec2}, \ref{atomfree}, and \ref{modularization} should make
clear that, for a sufficiently distorted array, it is enough that $g$ is strong enough to enable the formation of
the four bi-localized dressed states $\ket{\psi^{(\pm)}_{b\pm}}$ only.
In \fig \ref{fig7}, we show how the onset of such dressing benefits polaritonic transfer
as the CCA is progressively distorted for a fixed value 
of the atom-field coupling strength $g$.  
For the uniform array, i.e., $\eta\ug0$ [see \fig\ref{fig7}(a)]
the transmission has a poor efficiency.
As we have set $\omega_a\ug0$ (middle of the free field spectrum),
thus not matching any field normal mode, and because $g$ is small, 
the evolution is dominated by its free field
dynamics. 
Hence, the atomic component of the initial polariton is about frozen \cite{makin09, almeida13} while the photonic
component propagates freely along the array, bouncing back and forth, with the dynamics ruled mostly by the unbound modes.
The polaritonic transition amplitude significantly increases already by introducing a small amount of distortion [see \fig\ref{fig7}(b)]. 
Now, the bound bi-localized modes dominate the 
dynamics and the transition amplitude accordingly exhibits a periodic behaviour. 
A small contribution from the photonic unbound states, which results in short-time beatings, is yet present. 
Moreover, $g$ is still not much higher than $\delta \omega$, hence the dressing of the bi-localized modes is not maximum.
In \fig\ref{fig7}(c), we further distort the CCA in a way that the transition amplitude reaches considerably higher values.
As a consequence, the required transmission time grows since the array distortion causes the gap $\delta \omega$
to decrease. However, based on the modularization scheme introduced in Section
\ref{modularization}, this drawback can be got around. This is shown in \fig\ref{fig7}(d) where we consider a CCA split into 3 (5) 
weakly-connected modules each comprising 10 (6) cavities. Note that, compared with \fig\ref{fig7}(c), the time required to complete
the polaritonic-state transfer is considerably shortened while the maximum transition amplitude is about unaffected.

%

Regardless of the interaction regime (single-resonance or strong coupling), the crucial factor 
affecting the transfer fidelity is the end-to-end localization amplitude, i.e., the occurrence of
bi-localization either in the case of a standard staggered CCA or the modularized (partitioned) one.
The key ingredient is thus inducing the formation of bi-localized field normal modes and tuning the atoms
on resonance with those. 
The QST speed, however, can be managed 
by setting the appropriate regime and/or modularizing the CCA as in Section \ref{modularization}.

\section{\label{sec6}Conclusions}

In this work, we addressed the problem of transferring faithfully quantum states across a CCA.
We have shown that, while a staggered pattern of hopping rates offers shorter QST times with respect to a uniform pattern, a further significant reduction of the transfer time is achievable by imposing modularization on top of the staggered pattern. The modularization scheme yields up to three orders of magnitude shorter transfer times with respect to an unmodularized staggered array already for 20-site CCAs, while the gain increases for longer CCAs without affecting the performance in terms of QST fidelity.
%

To accomplish this task, we first focused on QST through a staggered atom-free CCA. By devising a 
perturbative approach to diagonalize analytically the Hamiltonian for a highly-distorted array, 
we showed that distortion induces the appearance of bound modes that are strongly bi-localized on the array
edges. In line with QST schemes exploiting bi-localization, this allows for high-fidelity QST.
As a distinctive property
of the staggered configuration, though, the scaling behaviour of the fidelity as a function of the CCA size has ideal features since,
for the high-distortion scenario, the fidelity is nearly insensitive to the array length (unlike in the case of a uniform bulk with weak outermost couplings). 
This yet
comes at the cost of having relatively long transfer times. To get around this drawback, we devised a strategy based on an engineered modularization
of the array into identical staggered subunits. We showed that in some paradigmatic instances this can result in a significant
reduction of the transfer time while maintaining the transfer fidelity about unchanged. Despite we focused on an atom-free
CCA, those findings apply to any spin chain regardless of the way 
it is implemented.

We then turned to a CCA where each cavity is coupled to an atom with the aim of exploring how the previous outcomes can be 
harnessed for transferring atomic or polaritonic states between the two array ends. In the weak-coupling regime where
the atomic frequency is resonant with one of the two bi-localized field modes, QST of atomic states can be achieved in a time set 
by the atom-field coupling strength. For stronger atom-photon couplings, one can instead exploit the formation of pairs of bi-localized dressed states
to efficiently transfer a polariton of given parity across the CCA 
in a time set by the energy gap between the pair of field bi-localized modes.

\section*{Acknowledgements}

This work was supported by CNPq. T.J.G.A. acknowledges the EU Collaborative Project TherMiQ (Grant No. 618074). F.C. acknowledges support
from Italian  PRIN-MIUR  2010/2011MIUR  (PRIN  2010-
2011). Fruitful discussions with D. L. Feder, S. Lorenzo and G. M. Palma are acknowledged.


%

\end{document}